# Basic Thermodynamics

*P. Duthil*[1]
Institut de Physique Nucléaire d'Orsay, IN2P3-CNRS/Université de Paris Sud, Orsay, France

**Abstract**
The goal of this paper is to present a general thermodynamic basis that is useable in the context of superconductivity and particle accelerators. The first part recalls the purpose of thermodynamics and summarizes its important concepts. Some applications, from cryogenics to magnetic systems, are covered. In the context of basic thermodynamics, only thermodynamic equilibrium is considered.

## 1 Introduction

Thermodynamics provides macroscopic descriptions of the states of complex systems and of their behaviours when they interact or are constrained under various circumstances. A *thermodynamic system* is a precisely specified macroscopic region of the Universe. It is limited by boundaries of a particular nature (real or not) with particular properties. All space in the Universe outside the thermodynamic system is known as the *surroundings*, the environment, or the reservoir. By (European) convention, the scalar quantities exchanged between the system and the surroundings are counted negatively if they leave the system and positively in the opposite case. A thermodynamically *isolated* system is completely isolated from its environment: it does not exchange energy (heat, work) or matter with the environment. If matter is able to flow in or out of its boundaries, the system is said to be *open*. Otherwise it is said to be *closed* (by real boundaries). Processes that affect the interior of the region of the system are studied using general principles: the *four laws of thermodynamics*. Although historically established before modern physics, most of these laws are based on the fact that a system is governed by general statistical principles which apply to the random motion of its particles. Hence, two approaches are possible when studying a system by consideration of its thermodynamics: a very general and practical one is to establish balances of the transfers between the system and its surroundings; the other is to apply statistical laws to the system. The latter approach is suitable for simple systems, and may also be used to clear up some of the more abstract concepts within thermodynamics.

A unique macroscopic description of a system requires thermodynamics to use a number of macroscopic parameters: the pressure $p$, the volume $V$, the magnetization $\hat{M}$, or the applied magnetic field $\hat{H}$, for example. Some other fundamental macroscopic parameters are defined by the laws of thermodynamics: the temperature $T$, the total internal energy $U$ and the entropy $S$. Knowing (experimentally) the minimum number of parameters required for this unique description fixes the number of independent state variables chosen for the macroscopic description. Thermodynamics then expresses constraints on the system by use of some relationships between these parameters. Parameters that are proportional to the size of the system are *extensive variables* and are written in the following in upper-case characters. Parameters that do not depend on the mass of the system are *intensive quantities* and are written in lower-case characters. An exception is made for the temperature, an intensive quantity denoted by $T$. For homogeneous systems, extensive parameters can be related to intensive variables via the relation $X = m \cdot x$, where $m$ is the mass. They then refer to bulk properties of the system. An example is the volume $V$ ($m^3$) and the *specific* volume $v$ ($m^3 \cdot kg^{-1}$).

[1] duthil@ipno.in2p3.fr

## 2 Equilibrium and equation of state

A thermodynamic system is in *thermodynamic equilibrium* when all its state variables remain constant with time: there is no net flow of matter or energy, no phase changes, and no unbalanced potentials (or driving forces) within the system. A system that is in thermodynamic equilibrium experiences no change when it is isolated from its surroundings. It should be stressed that thermodynamic equilibrium implies steady state, but that steady state does not always induce thermodynamic equilibrium (e.g. steady heat diffusion in a support). For different combined subsystems forming one thermodynamic system at thermodynamic equilibrium, the extensive parameters of each subsystem are summed (volume, for example) and the intensive parameters are equal (pressure, temperature). At equilibrium, a functional relationship between the parameters (state variables) of a system can be expressed as an *equation of state*.

As an example, a gas is completely described at equilibrium by the parameters $p$, $V$, and $T$. The equation of state, of the form $f(p,V,T) = 0$, describes an equilibrium surface on which a point represents an equilibrium state. The isotherms of a (dilute) gas may be described according to Boyle's law (or the Boyle–Mariotte law) by $pV/n = cst$, where $n$ denotes the number of moles. An experimental temperature scale can be defined by any function of the temperature associated with these isotherms. The simplest function is $pV = nRT$, known as the *ideal gas law*, which defines the isotherms as hyperbolic curves in the $p-V$ plane—see Fig. 1(a). Using an air thermometer, Amontons suggested in 1702 that the zero on this temperature scale would be for the zero pressure limit. Taking into account the interaction between the molecules, the van der Waals equation $(p + a/v^2)(v - b) = rT$ is also an equation of state for a gas: it modifies the pressure term by considering the attraction effect $-a$ between molecules; it also corrects the volume by considering what is really accessible by one mole of molecules: a volume $b$ is indeed not available due to the repulsion effect of the molecules. It is accurately applicable for a larger range of pressures. In the van der Waals equation, there is a critical isotherm for which $(\partial p/\partial V)_T = (\partial^2 p/\partial V^2)_T = 0$ at a *critical point* C having coordinates $p_C = a/(27b^2)$, $V_C = 3nb$, and $T_C = 8a/(27Rb)$. Tending to this critical point, the van der Waals hyperbolic isotherms experience an inflection point—see Fig. 1(b). Below this critical point, two inflection points are revealed (furthermore, for a given volume and temperature, the calculated pressure can be negative). This domain coincides with the liquid phase of the fluid (see Section 15).

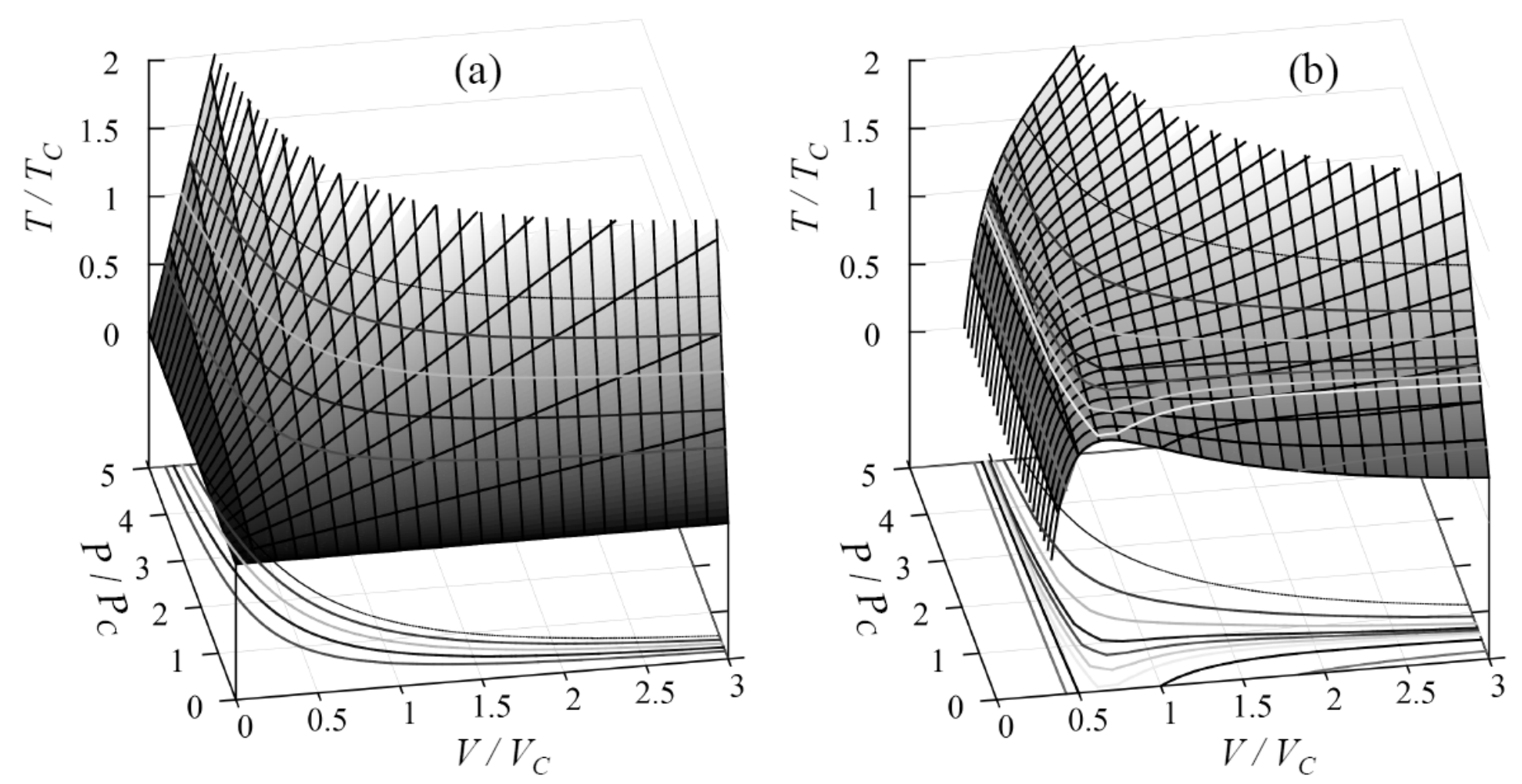

**Fig. 1**: Equilibrium surfaces given by the ideal gas (a) and the van der Waals (b) equations of state

## 3 Transformations

Thermodynamics often consider a system evolving from an initial equilibrium state to a final equilibrium state via a *thermodynamic transformation*. A *reversible* transformation is a thermodynamic process that can be assessed via a succession of thermodynamic equilibria by infinitesimally modifying some external constraints. It can be reversed without changing the nature of the external constraints, and only by reversing the parameters (such as time). Otherwise it is *irreversible*. When the thermodynamic process occurs infinitely slowly, it might ensure that the system goes through a sequence of states that are infinitesimally close to equilibrium: state variables and equation of states are well defined. But the system is not in equilibrium with its surroundings: the transformation is thus *quasi-static*. A reversible transformation is quasi-static. But a quasi-static transformation can be irreversible (heat diffusion in a support, for example). If the final state is the same as the initial state, the process is a *thermodynamic cycle*.

## 4 Work, heat, and the first law of thermodynamics

### 4.1 Work

Mechanical work $\delta W = F \cdot \mathrm{d}x$ is achieved when displacement d$x$ or deformation occurs by means of a force field. Consider a closed system, for example a gas contained in a cylinder, as shown in Fig. 2(a).

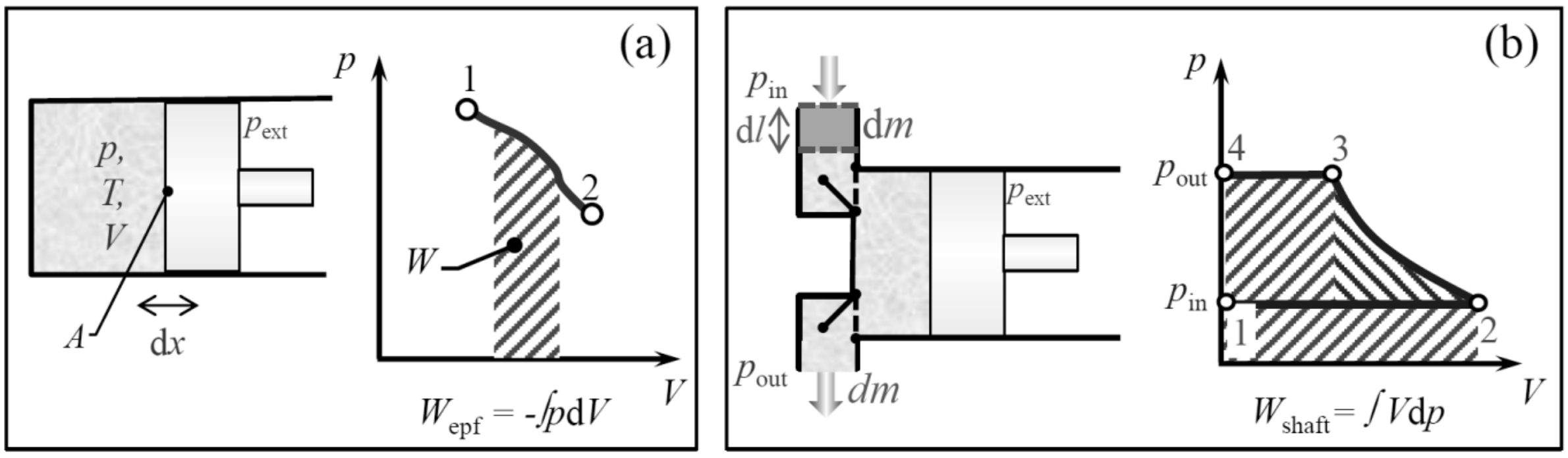

**Fig. 2:** Work of a closed system (a) and an open system (b)

For a reversible expansion or compression of the gas, induced by the movement of a piston of cross-sectional area $A$,

$$\delta W_{\text{epf}} = -F \ \mathrm{d}x = -F / A \cdot A \ \mathrm{d}x = -p \ \mathrm{d}V$$

and

$$W_{\text{epf}} = \int -p \ \mathrm{d}V \ , \tag{1}$$

where the subscript ‘epf’ refers to the external pressure forces. During expansion, the volume change $\mathrm{d}V$ is positive and thus $\delta W_{\text{epf}} < 0$: work is given by the system to the surroundings. During compression, work is received from the surroundings. For an *isochoric process*, $\mathrm{d}V = 0$ and thus $\delta W_{\text{epf}} = 0$.

Considering the open system shown in Fig. 2(b), a mass of gas d$m$ is received by the cylinder during an admission step preceding the expansion and is then rejected after compression. The flow of matter entering the system achieves some work as a piston of fluid; the system also performs work onto the exiting flow of matter. This total flow work is given by:

$$\delta W_{\text{flow}} = p_{\text{in}} A_{\text{in}} \ \mathrm{d}l_{\text{in}} - p_{\text{out}} A_{\text{out}} \ \mathrm{d}l_{\text{out}} = p_{\text{in}} \ \mathrm{d}V_{\text{in}} - p_{\text{out}} \ \mathrm{d}V_{\text{out}} = p_{\text{in}} v_{\text{in}} \delta m_{\text{in}} - p_{\text{out}} v_{\text{out}} \delta m_{\text{out}} = \left[ p v \delta m \right]_{\text{out}}^{\text{in}} .$$

where $v$ is the specific volume (unit: $m^3 \cdot kg^{-1}$) of the fluid. The work which may be performed by the system on some mechanical device is often called the 'shaft' work (useful work), and is the difference between the total work $\delta W$ and the flow work $\delta W_{\text{flow}}$:

$$\delta W_{\text{shaft}} = \delta W - \delta W_{\text{flow}} = V \, \mathrm{d}p \; . \tag{2}$$

### 4.2 Internal energy

The *internal energy* $U$ (unit: J) is the sum of the kinetic and potential energies of all the physical particles of the system (and of the rest-mass energies) at some microscale: $U = \sum E_{\text{c,micro}} + \sum E_{\text{p,micro}}$. It is a state function and can thus be defined by macroscopic parameters. For example, for a non-magnetic compressible fluid, if $p$ and $V$ are fixed, $U = U(p,V)$ is also fixed.

### 4.3 First law of thermodynamics and energy balances

Between two thermodynamic equilibria, $\delta U = \delta W + \delta Q$; for a reversible process, $\mathrm{d}U = \delta W + \delta Q$, where $W$ is the exchange work (mechanical, electrical, magnetic interaction, etc.) and $Q$ is the exchange heat. For a cyclic process during which the system evolves from an initial state $I$ to an identical final state $F$: $\Delta U = U_F - U_I = 0$.

Between two thermodynamic equilibria, the total energy change is given by $\Delta E = \Delta E_{\text{c,macro}} + \Delta E_{\text{p,macro}} + \Delta U = W + Q$. $E_{\text{c,macro}}$ and $E_{\text{p,macro}}$ are the kinetic and potential energies of the system at a macroscopic scale, and they are null if the system does not move or is not in a force field. This leads to:

$$\Delta U = W + Q \, . \tag{3}$$

If work is due to external pressure forces only, then $W = W_{\text{epf}} = \int -p \, \mathrm{d}V$ and $\Delta U = W_{\text{epf}} + Q$. For an isochoric process, the volume remains constant and $\Delta U = Q$. This relation is the basis of calorimetry.

For an open system, the work $\delta W_{\text{flow}}$ achieved by the pressure forces to make the fluid circulate in and out is distinguished from the useful work performed by other forces (mechanical, magnetic, electrical, etc.) and denoted $\delta W_{\text{u}}$. We note that for the system of Fig. 2(b) $\delta W_{\text{u}} = \delta W_{\text{shaft}}$. The first law then yields:

$$\Delta E = \Delta E_{\text{c,macro}} + \Delta E_{\text{p,macro}} + \Delta U + \delta m \left[u\right]_{\text{inlet}}^{\text{outlet}} = W_{\text{u}} + W_{\text{flow}} + Q \, .$$

This can be rewritten as follows:

$$\Delta E_{\text{c,macro}} + \Delta E_{\text{p,macro}} + \Delta U + \delta m \left[u + pV\right]_{\text{inlet}}^{\text{outlet}} = W_{\text{u}} + Q \, .$$

Another function of state, called *enthalpy*, is then defined as

$$H = U + pV \quad \text{(unit: J)}. \tag{4}$$

Thus, $h$ being the enthalpy per unit mass ($J \cdot kg^{-1}$), we obtain

$$\Delta E = \Delta E_{\text{c,macro}} + \Delta E_{\text{p,macro}} + \Delta U + \delta m \left[h\right]_{\text{inlet}}^{\text{outlet}} = W_{\text{u}} + Q \, .$$

For a cyclic process ($\Delta U = 0$), if $\Delta E_{\text{c,macro}} = \Delta E_{\text{p,macro}} = 0$,

$$\Delta H = W_{\text{u}} + Q \; . \tag{5}$$

In the case of an isobaric process (for which the pressure remains constant) and if no other type of work is performed other than the pressure work, $\Delta H = Q$.

The energy balance for an open system at steady state or undergoing a cyclic process is written as follows:

$$W_u + Q = \sum_{\text{outlet boundaries}} mh - \sum_{\text{inlet boundaries}} mh \quad \text{(unit: J)},$$
$$\dot{W}_u + \dot{Q} = \sum_{\text{outlet boundaries}} \dot{m}h - \sum_{\text{inlet boundaries}} \dot{m}h \quad \text{(unit: W)}. \tag{6}$$

## 5 Entropy and second law of thermodynamics

For a system considered to be between two successive states, there is a function of state $S$ called *entropy*, unit $J \cdot K^{-1}$, such that:

$$\Delta S_{syst} = \Delta S = \Delta S^e + S^i , \tag{7}$$

where $\Delta S^e = \int \delta Q / T$ relates to the heat exchange and $S^i$ is an entropy production term: $S^i = \Delta S_{syst} + \Delta S_{surroundings}$. For a reversible process, $S^i = 0$; for an irreversible process, $S^i > 0$. For an *adiabatic* ($\delta Q = 0$) and reversible process ($S^i = 0$) the thermodynamic transformation is *isentropic*: $\Delta S = 0$.

### 5.1 Entropy of an isolated system

An isolated system has no exchange with the surroundings; thus $\Delta S^e = 0$ and $\Delta S_{syst} = S^i \geq 0$. At thermodynamic equilibrium, the isolated system experiences no change, and its state variables remain constant with time. Hence, $\Delta S_{syst} = 0$ and $S^i = 0$.

The statistical weight $\Omega$ denotes the number of possible microstates of a system: the position of the atoms or molecules, or the distribution of the internal energy. The different possible microstates correspond to (and are consistent with) the same macrostate described by the macroscopic parameters ($p$, $V$, etc.). The probability of finding the system in one microstate is the same as that of finding it in another microstate. Thus, the probability that the system is in a given macrostate must be proportional to $\Omega$. The entropy can be expressed as follows:

$$S = k_B \ln(\Omega) , \tag{8}$$

where $k_B = 1.38 \cdot 10^{-23}$ $J \cdot K^{-1}$ is the Boltzmann constant. Considering two systems $A$ and $B$, the number of microstates of the combined system $A \cup B$ is $\Omega_A \cdot \Omega_B$ and its entropy is $S = S_A + S_B$: the entropy is additive. Similarly, it is proportional to the mass of the system, and is thus an extensive parameter, i.e. $S = m \cdot s$, where $s$ is the specific entropy (unit: $J \cdot K^{-1} \cdot kg^{-1}$).

### 5.2 Principle of increase of entropy

One statement of the second law of thermodynamics is that “entropy of an isolated system tends to increase”. The system changes in response to this tendency. Such a change is said to be ‘spontaneous’. If one imagines two possible macrostates for an isolated system, corresponding to two values of the number of possible microstates $\Omega$, the system will experimentally be found in the macrostate for which the number of microstates is the greater, thus having larger entropy. We consider, for example, a gas in a box, and assume that the gas is on one side of the box at some initial time corresponding to $\Omega = \Omega_1$. If the gas is free to fill up the whole box, corresponding to $\Omega = \Omega_2$, it will ‘spontaneously’ go to that state. Indeed, $\Omega_2 >> \Omega_1$, and there is a higher probability of finding the gas throughout the whole box than of staying on one side of the box. This principle applies to isolated systems only. Otherwise, one would have to consider exchanges with the surroundings, and thus the change of entropy of the surroundings. Even though not very convenient, it is always possible to transform a system into an isolated one by extending its boundaries, keeping in mind that the whole Universe is an isolated system.

## 6 Thermodynamic temperature and the zeroth law of thermodynamics

Let us consider two closed systems, $A$ at temperature $T_A$ and $B$ at $T_B$, each having a constant volume and separated by a diathermal wall: they are in thermal contact, they are not isolated from each other, but they cannot exchange work (or matter). The combined system $A \cup B$ is isolated and therefore $\mathrm{d}(U_A + U_B) = 0 = \delta Q_A + \delta Q_B$. Thus, energy $\delta U = \delta Q$ (heat) can flow from $A$ to $B$ and from $B$ to $A$ such that $\delta Q_A = -\delta Q_B$. The net change of entropy of the isolated combined system is given by

$$\delta S = \delta S^{\mathrm{e}} + \delta S^{\mathrm{i}} = 0 + \delta S^{\mathrm{i}} = \delta S_A + \delta S_B = \left( \frac{1}{T_A} - \frac{1}{T_B} \right) \delta Q_A \geq 0 \ .$$

Indeed, if $T_A > T_B$, we expect that heat will flow spontaneously from $A$ to $B$ such that $\delta Q_A < 0$ and then $\delta S > 0$. If, however, $T_A < T_B$, then $\delta Q_B = -\delta Q_A < 0$ and $\delta S > 0$. We find a relationship between temperature and entropy which is consistent with the second law of thermodynamics.

The *thermodynamic temperature* is defined as:

$$T = \left( \partial U / \partial S \right)_V . \tag{9}$$

For the isolated combined system $A \cup B$, the net change of entropy can rewritten as

$$\delta S = \left( \frac{\partial S_A}{\partial U_A} \right)_{V_A} \delta U_A + \left( \frac{\partial S_B}{\partial U_B} \right)_{V_B} \delta U_B = \left( \frac{1}{T_A} - \frac{1}{T_B} \right) \delta Q_A .$$

At thermodynamic equilibrium, maximization of the entropy requires that $\delta S = \delta S^{\mathrm{e}} + S^{\mathrm{i}} = 0 + 0 = 0$ and thus that $T_A = T_B$. Similarly, one can observe that two systems both in thermal equilibrium with a third system are in thermal equilibrium with each other. This equilibrium can be assessed by the equality of scalar quantities, i.e. the temperature of each system, an intensive parameter. This is the zeroth principle of thermodynamics, or the principle of temperature existence.

## 7 Boltzmann factor and the third law of thermodynamics: statistical point of view

We now consider a system $SYST$, having energy $E$, in relation to its much larger surroundings or environment, named $EXT$, which has energy $E_{EXT}$ at temperature $T$. We assume that the combined system $TOT = SYST \cup EXT$ is isolated (system $E_{EXT}$ is possibly the entire Universe outside $SYST$). Then the probability that $SYST$ has energy $E$ (corresponding to the number of microstates $\Omega$) is the probability that $EXT$ has energy $E_{TOT} - E$ corresponding to the number of microstates $\Omega_{EXT}\left(E_{TOT} - E\right)$. Noting that $S_{EXT} = k_{\mathrm{B}} \ln \Omega_{EXT}$, we have $\ln \Omega_{EXT}\left(E_{TOT} - E\right) = S_{EXT} / k_{\mathrm{B}}$. $S_{EXT}$ is thus a function of $E_{TOT} - E = E_{EXT}$, which can be expanded around $E_{TOT}$ (noting that $E << E_{TOT}$):

$$\ln \Omega_{EXT}\left(E_{TOT} - E\right) = \frac{1}{k_{\mathrm{B}}} S_{EXT}(E_{TOT}) - \frac{E}{k_{\mathrm{B}}} \left( \frac{\partial S_{EXT}}{\partial E_{TOT}} \right)_{E=0} = \frac{1}{k_{\mathrm{B}}} S_{EXT}(E_{TOT}) - \frac{E}{k_{\mathrm{B}} T} = \ln \Omega_{EXT}(E_{TOT}) - \frac{E}{k_{\mathrm{B}} T} \ .$$

The probability of observing the macroscopic system $SYST$ with energy $E$ is thus proportional to the Boltzmann factor: $\exp(-E / k_{\mathrm{B}} T)$. The temperature limit $T \to 0$ means that the system is in a minimum-energy state. At the zero of temperature, any system in thermal equilibrium must exist in its lowest possible energy state. If we consider, for example, $SYST$ to be a molecule with energy $\varepsilon$ in an ideal gas of constant volume and energy $E_{EXT}$, the probability of the molecule having energy $\varepsilon$ is $\exp(-\varepsilon / k_{\mathrm{B}} T)$. As $T \to 0$, the whole gas is in a state of minimum energy as no molecule can be in an excited state. Thus, if $T \to 0$, the minimum-energy state is unique ($\Omega \to 1$) and $S = 0$. It therefore follows that an absolute entropy can be computed and referenced from the temperature.

## 8 Relating entropy to functions of state

### 8.1 Entropy with internal energy

$U$ and $S$ are state functions. Any macrostate at equilibrium can be described by a relationship between these functions and the independent state variables. For example, a simple compressible system is completely characterized by its energy and its volume, and $U = f(S,V)$. Therefore, for a reversible process ($S^{\mathrm{i}} = 0$), the change of the state functions can be written as

$$\mathrm{d}U = \left(\frac{\partial U}{\partial S}\right)_V \mathrm{d}S + \left(\frac{\partial U}{\partial V}\right)_S \mathrm{d}V = T\ \mathrm{d}S + \left(\frac{\partial U}{\partial V}\right)_S \mathrm{d}V\ .$$

Identifying this equation with the expression of the first law of thermodynamics for a closed system yields $p = -(\partial U / \partial V)_S$ and thus

$$\mathrm{d}U = T\ \mathrm{d}S - p\ \mathrm{d}V\ . \tag{10}$$

The kinetic temperature of an ideal gas can be derived from the expression of its internal energy and from the motion of its molecules. This kinetic temperature can be related to the thermodynamic temperature using the Boltzmann distribution. It follows that the equation of state of an ideal gas can be written as

$$pV = Nk_{\mathrm{B}}T = nN_{\mathrm{A}}k_{\mathrm{B}}T\ , \tag{11}$$

where $N$ is the number of molecules of the gas and $N_{\mathrm{A}} = 6.022 \cdot 10^{23}\ \mathrm{mol}^{-1}$ is Avogadro's number. Stating that $R = N_{\mathrm{A}}k_{\mathrm{B}}$, we find again $pV = nRT$, and thus we relate the experimental temperature defined in Section 2 to both the thermodynamic and kinetic temperature.

### 8.2 Thermodynamic potentials

We consider a non-isolated system in thermal contact with its surroundings ($EXT$) which is at temperature $T_{EXT}$, with a large heat capacity such that heat flows from the surroundings into the system without changing $T_{EXT}$. The second principle states that for the system combined with its surroundings the total change of entropy is given by $\delta S_{TOT} = \delta S + \delta S_{EXT} \geq 0$ with $\delta S_{EXT} = -\delta Q / T_{EXT}$. For the considered system having a constant volume, $\delta U = \delta Q$. Thus, we obtain $\delta(U - T_{EXT}S) \leq 0$: this spontaneous change is related to a decrease of $U - T_{EXT}S$.

A new thermodynamic equilibrium ($T = T_{EXT}$) will thus be reached for the minimum of the *(Helmholtz) free energy* of the system $F = U - TS$. Similarly, for a system in thermal contact with its surroundings and held at constant pressure, the *Gibbs free energy* $G = U - TS + pV$ tends to a minimum at thermodynamic equilibrium. The free energy and the Gibbs free energy are *thermodynamic potentials*.

Thermodynamic potentials are specific functions of some state variables: the natural variables of the potential. When these natural variables are kept constant, a closed thermodynamic system evolves towards equilibrium if the potentials diminish. The thermodynamic equilibrium corresponds to a minimum of one of the thermodynamic potentials. The internal energy $U = f(S,V)$ is such a thermodynamic potential of natural variables $S$ and $V$. For a mechanical system (no heat exchange), we saw that an increase of energy is the result of a gain of mechanical work which is the product of a force by a small displacement. We can generalize this concept by stating that the increase in energy of a thermodynamic system is the result of the addition of several products of generalized thermodynamic forces which induce thermodynamic displacements from equilibrium. The thermodynamic force is driven by an intensive variable which is conjugated to an extensive variable representing the thermodynamic displacement: $p$ and $V$ are such conjugated variables, as are $T$ and $S$. If $S$ and $V$ are kept constant, $U$ will tend to a minimum as the considered closed system tends to thermodynamic equilibrium. The other thermodynamic potentials tending to a minimum for a closed system at equilibrium are:

the enthalpy, if $p$ and the extensive parameters (such as $S$) are held constant,

$$H = U + pV \Rightarrow \mathrm{d}H = T\ \mathrm{d}S + V\ \mathrm{d}p\,; \tag{12}$$

the Helmholtz free energy, if $T$ and the extensive parameters (such as $V$) are held constant,

$$F = U - TS \Rightarrow \mathrm{d}F = -S\ \mathrm{d}T - p\ \mathrm{d}V\,; \tag{13}$$

the Gibbs free energy if $p$, $T$, and the extensive parameters are held constant,

$$G = U - TS + pV \Rightarrow \mathrm{d}G = -S\ \mathrm{d}T + V\ \mathrm{d}p\,. \tag{14}$$

## 9 Maxwell relations

For a compressible system $U = f(S,V)$, the differential form $\mathrm{d}U$ is exact if the two mixed second partial derivatives are equal:

$$\frac{\partial^2 U}{\partial V \partial S} = \frac{\partial^2 U}{\partial S \partial V} \Rightarrow \left(\frac{\partial}{\partial V}\right)_S \left(\frac{\partial U}{\partial S}\right)_V = \left(\frac{\partial}{\partial S}\right)_V \left(\frac{\partial U}{\partial V}\right)_S .$$

As $\mathrm{d}U = T\ \mathrm{d}S - p\ \mathrm{d}V$, we deduce an appropriate relation between the first derivatives of $T$ and $p$:

$$\left(\frac{\partial T}{\partial V}\right)_S = -\left(\frac{\partial p}{\partial S}\right)_V .$$

By the same argument, we obtain from the different potentials the useful Maxwell relations:

$$\begin{aligned} U = U(S,V) &\Rightarrow \left(\frac{\partial T}{\partial V}\right)_S = -\left(\frac{\partial p}{\partial S}\right)_V, \qquad & F = F(T,V) &\Rightarrow \left(\frac{\partial p}{\partial T}\right)_V = \left(\frac{\partial S}{\partial V}\right)_T, \\ H = H(S,p) &\Rightarrow \left(\frac{\partial T}{\partial p}\right)_S = \left(\frac{\partial V}{\partial S}\right)_p, \qquad & G = G(T,p) &\Rightarrow \left(\frac{\partial S}{\partial p}\right)_T = -\left(\frac{\partial V}{\partial T}\right)_p . \end{aligned} \tag{15}$$

## 10 Heat capacities

The amount of heat that must be added to a system *reversibly* to change its temperature is the heat capacity: $C = \delta Q / dT$ (unit: $\mathrm{J \cdot K^{-1}}$). The conditions under which heat is supplied, at constant volume $V$ or constant pressure $p$, must be specified:

$$C_V = T\left(\frac{\partial S}{\partial T}\right)_V = \left(\frac{\partial U}{\partial T}\right)_V, \quad C_p = T\left(\frac{\partial S}{\partial T}\right)_p = \left(\frac{\partial H}{\partial T}\right)_p . \tag{16}$$

It turns out that

$$\text{for } V = cst,\ S(T,V) = \int_0^T \frac{C_V}{T}\ \mathrm{d}T \text{ and } \Delta U = \int C_V\ \mathrm{d}T\,;$$

$$\text{for } p = cst,\ S(T,p) = \int_0^T \frac{C_p}{T}\ \mathrm{d}T \text{ and } \Delta H = \int C_p\ \mathrm{d}T\,.$$

$Q = \int C_i dT$ is known as the 'sensible heat' (see the lecture 'Materials properties at low temperature' in these proceedings).

The relationship between $C_p$ and $C_V$ is expressed by Meyer's relation:

$$C_p - C_V = T\left(\frac{\partial p}{\partial T}\right)_V \left(\frac{\partial V}{\partial T}\right)_p , \tag{17}$$

which yields, for an ideal gas,

$$C_p - C_V = nR, \; C_V = n\frac{R}{\gamma-1}, \text{ and } C_p = n\frac{\gamma R}{\gamma-1} , \tag{18}$$

where $\gamma$ is the ratio of the heat capacities:

$$\gamma = \frac{C_p}{C_V} = T\left(\frac{\partial V}{\partial p}\right)_T = \left(\frac{\partial V}{\partial p}\right)_S^{-1} . \tag{19}$$

## 11 Some thermodynamic transformations

### 11.1 Adiabatic expansion of a gas

For a reversible adiabatic expansion,

$$\left(\frac{\partial T}{\partial p}\right)_S = -\left(\frac{\partial T}{\partial S}\right)_p \left(\frac{\partial S}{\partial p}\right)_T = \left(\frac{\partial T}{\partial S}\right)_p \left(\frac{\partial V}{\partial T}\right)_p = \frac{T}{C_p}\left(\frac{\partial V}{\partial T}\right)_p .$$

As $C_p > 0$ and $(\partial V/\partial T)_p > 0$, $(\partial T/\partial p)_S > 0$. Thus, as for an expansion $\partial p < 0$, we get $\partial T < 0$, expressing a cooling. During the adiabatic expansion of a gas, work is extracted and the gas is cooled. This is achieved in a reciprocating engine or in a turbine (turbo-expander).

### 11.2 Joule–Kelvin (Joule–Thomson) expansion of a gas

We consider a gas flowing and expanding through a throttling valve from a fixed high pressure $p_1$ to a fixed low pressure $p_2$. The (open) system is thermally isolated and thus the transformation is *isenthalpic*:

$$\Delta U = W \Rightarrow U_1 + p_1 V_1 = U_2 + p_2 V_2 \Rightarrow H_1 = H_2 .$$

It follows that:

$$\left(\frac{\partial T}{\partial p}\right)_H = -\left(\frac{\partial T}{\partial H}\right)_p \left(\frac{\partial H}{\partial p}\right)_T = -\frac{1}{C_p}\left(\frac{\partial H}{\partial p}\right)_T = -\frac{1}{C_p}\left[\left(\frac{\partial S}{\partial p}\right)_T + V\right] = \frac{V}{C_p}(\alpha T - 1) ,$$

where $\alpha$ is the coefficient of thermal expansion:

$$\alpha = \frac{1}{V}\left(\frac{\partial V}{\partial T}\right)_p . \tag{20}$$

For the ideal gas, $\alpha T = 1$ and thus $(\partial T/\partial p)_H = 0$: the isenthalpic expansion does not change the temperature of the gas.

For a real gas, if $\alpha T < 1$, $(\partial T/\partial p)_H > 0$, and the expanding gas is warmed; if $\alpha T > 1$, $(\partial T/\partial p)_H < 0$, and the gas is cooled. The latter case occurs when the gas temperature at the inlet of the valve is below a certain temperature called the *inversion temperature*. The inversion temperature curve of helium-4 ($^4$He) is plotted as a function of pressure and temperature in Fig. 3. The maximum inversion temperature (as $p \to 0$) is about 45 K. In a helium liquefier (or refrigerator), the gas is usually subcooled below the inversion temperature by combining adiabatic expansion and heat removal within turbines and heat exchangers. It is finally liquefied by use of a Joule–Thomson expansion. Note that the maximum inversion temperatures of hydrogen, nitrogen and oxygen are 202 K, 623 K and 761 K, respectively.

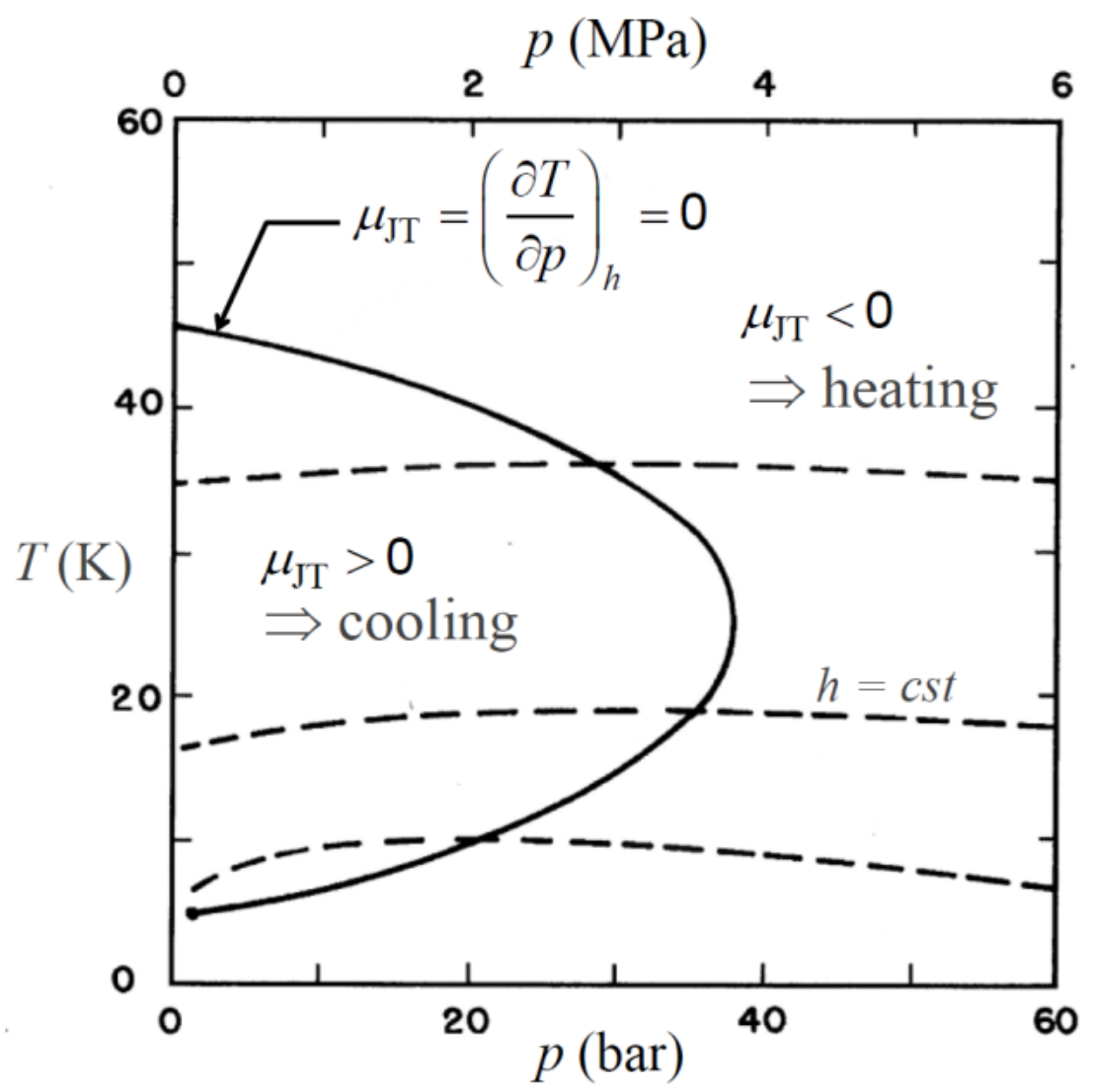

**Fig. 3:** Inversion temperature of helium-4 as a function of $p$ and $T$. Constant enthalpy lines are dashed; the solid line is the inversion curve.

### 11.3 Thermodynamic transforms of an ideal gas

Table A.1 of Appendix A lists different reversible thermodynamic processes involving an ideal gas. It provides a plot of the considered evolution in a $p-V$ diagram, expresses the heat and work exchanged, and gives the changes in internal energy, enthalpy and entropy.

## 12 Heat machines

The basic function of a heat machine is to convert heat into work or, conversely, to produce a heat transfer by means of supplied work. This work can be electrical, mechanical, etc., but the second principle states that its production by a *heat engine* requires at least two heat reservoirs at two different temperatures: the machine is driven with some heat supplied by the hotter reservoir, produces work, and dumps waste heat to a colder reservoir (see Fig. 4). A *heat pump* is driven by an amount of work and aims to transfer heat from a colder reservoir to a hotter one. It can therefore be used to warm or to cool. In the latter case, the heat pump is a *refrigerator*. A dithermal machine is a heat machine working by way of two heat reservoirs: one hot at temperature $T_{\mathrm{H}}$ and one cold at $T_{\mathrm{C}}$. Over one thermodynamic cycle, the system in the final state recovers its initial states. Thus,

$$\text{energy balance (first law): } \Delta U = U_{\mathrm{f}} - U_{\mathrm{i}} = 0 \quad \Rightarrow \quad W + Q_{\mathrm{C}} + Q_{\mathrm{H}} = 0\,;$$

$$\text{entropy balance (second law): } \Delta S = \Delta S^{\mathrm{e}} + S^{\mathrm{i}} = 0 \quad \Rightarrow \quad Q_{\mathrm{C}} / T_{\mathrm{C}} + Q_{\mathrm{H}} / T_{\mathrm{H}} + S^{\mathrm{i}} = 0\,.$$

It follows that:

$$Q_{\mathrm{H}} = -Q_{\mathrm{C}} - \mathrm{W} = -\mathrm{T}_{\mathrm{H}} / T_{\mathrm{C}} \cdot Q_{\mathrm{C}} - T_{\mathrm{H}} S^{\mathrm{i}}\;. \tag{21}$$

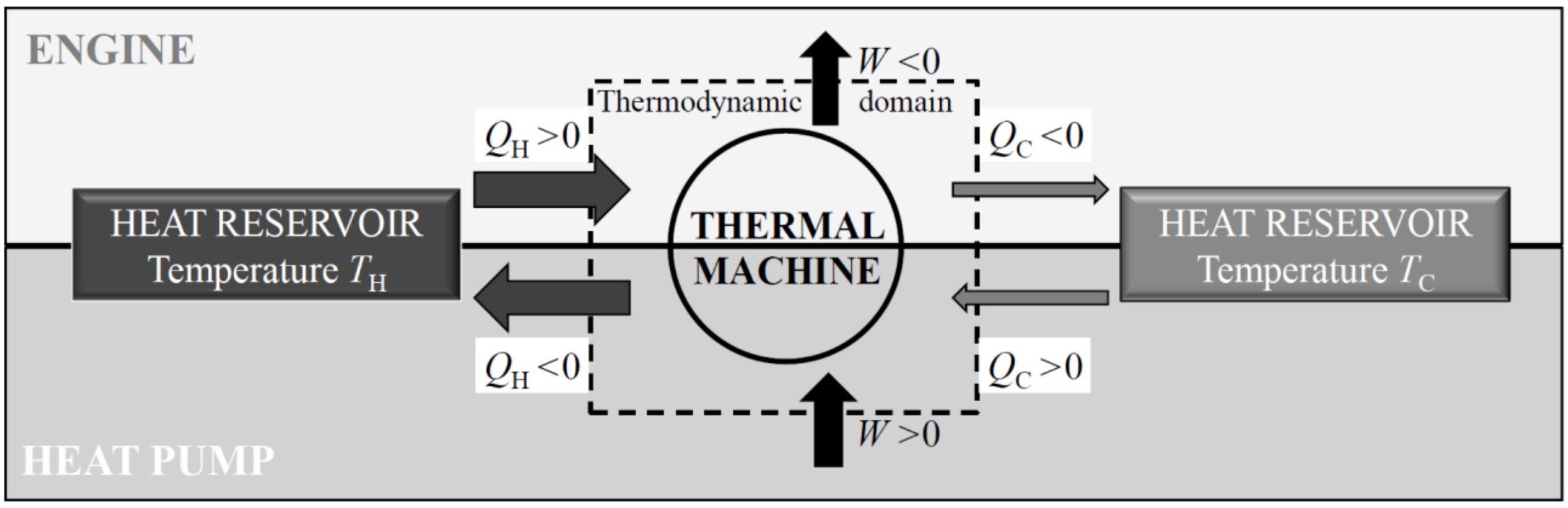

**Fig. 4:** Principle of a heat machine: engine (top) and heat pump or refrigerator (bottom)

From the two expressions given for the energy and entropy balances, we can plot the domain of the existence of heat machines in a $Q_H - Q_C$ diagram, called a Raveau diagram and shown in Figs. 5(a) and (b).

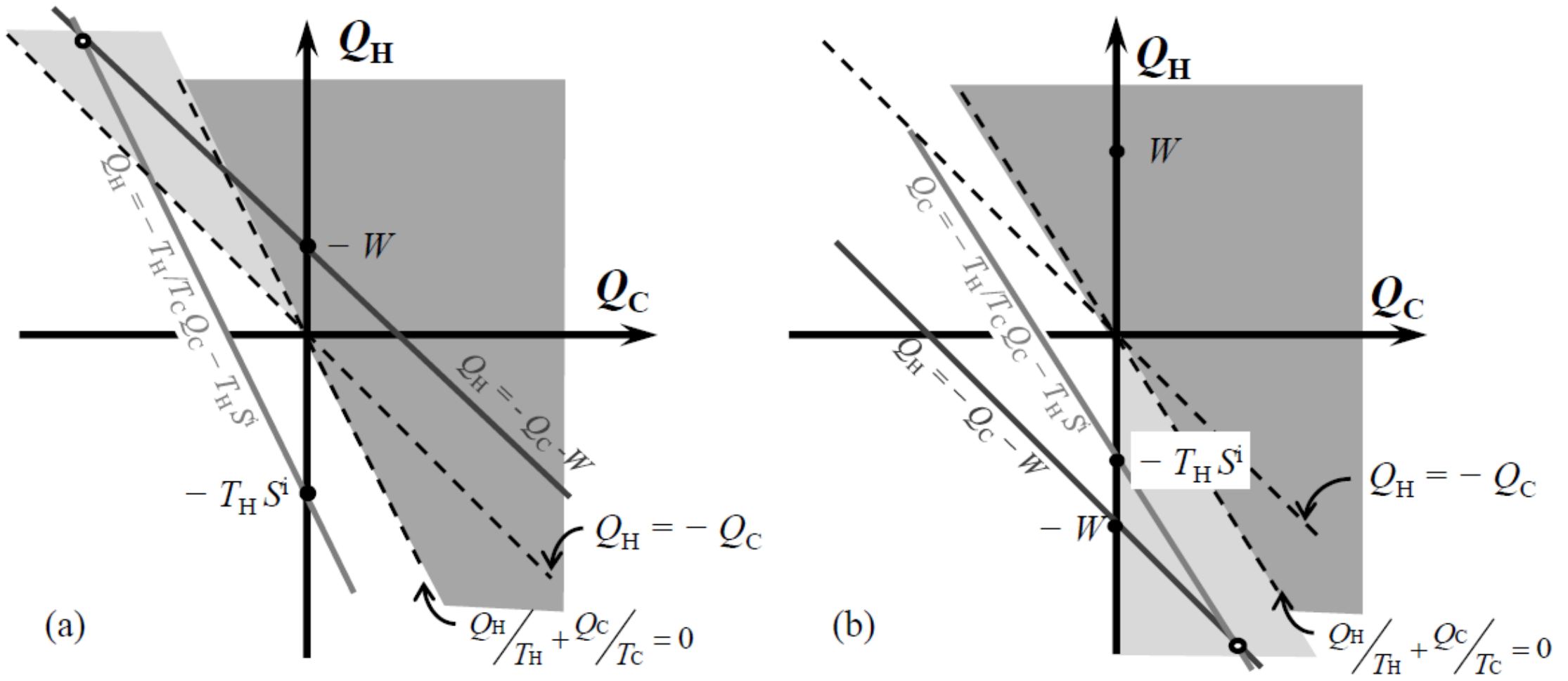

**Fig. 5:** Raveau diagrams of a heat engine (a) and a heat pump (b). Domains of existence of both types of heat machines are indicated by pale grey zones. The domains of impossible cycles (as in not respecting the second law) are indicated in dark grey.

## 12.1 Heat engine

As the work is produced (given) by the engine, we have $W<0$ and, if $T_H > T_C$, then $Q_H > 0$ and $Q_C < 0$ (see Fig. 5(a)). The engine is supplied by heat coming from the hot reservoir and rejects excess heat at the cold reservoir. The *efficiency* $\eta$ of an engine can be defined by the (positive) ratio of the amount of produced work to the energy that is supplied. The supply being the heat coming from the hot source, we have

$$\eta = \frac{|W|}{Q_H} = \frac{-W}{Q_H} = \frac{Q_H + Q_C}{Q_H} = 1 + \frac{Q_C}{Q_H},$$

i.e., making use of equation (21),

$$\eta = 1 - T_C / T_H - T_C S^i / Q_H \le \eta_C = 1 - T_C / T_H < 1 \text{ since } S^i \ge 0. \quad (22)$$

The maximum possible efficiency $\eta_C$ is obtained in the case of no irreversibility ($S^i = 0$) and is called the *Carnot efficiency*, referring to the ideal Carnot cycle presented in Section 12.3.

## 12.2 Heat pump or refrigerator

Similarly, for a heat pump or a refrigerator, work is supplied: $W > 0$. Thus, if $T_H > T_C$, then $Q_H < 0$ and $Q_C > 0$ (see Fig. 5(b)); heat is pumped from the cold reservoir and heat is rejected at the hot reservoir. The efficiency of a heat pump or a refrigerator is defined by the Coefficient Of Performance (*COP*), which is the ratio of the energy of interest to the energy supplied to the machine. The energy of interest has to be distinguished between a heat pump, which is designed to warm, and a refrigerator, which is designed to cool.

Thus, for a heat pump,

$$COP_{\text{Heating}} = \frac{|Q_H|}{W} = \frac{-Q_H}{W} = \frac{-Q_H}{-Q_H - Q_C} = \frac{1}{1 + Q_C / Q_H} ,$$

and, noting that $COP_{\text{C Heating}}$ is the coefficient of performance of an ideal Carnot heat pump, we have

$$COP_{\text{Heating}} = \frac{1}{1 - T_C / T_H - T_C S^{i} / Q_H} \leq COP_{\text{C Heating}} = \frac{1}{1 - T_C / T_H} . \tag{23}$$

For a refrigerator,

$$COP_{\text{Cooling}} = \frac{Q_F}{W} = \frac{Q_F}{-Q_H - Q_C} = \frac{1}{-1 - Q_H / Q_C} ,$$

and, noting that $COP_{\text{C Cooling}}$ is the coefficient of performance of an ideal Carnot heat pump, we have

$$COP_{\text{Cooling}} = \frac{1}{T_H / T_C - 1 + T_C S^{i} / Q_H} \leq COP_{\text{C Cooling}} = \frac{1}{T_H / T_C - 1} . \tag{25}$$

## 12.3 The Carnot cycle

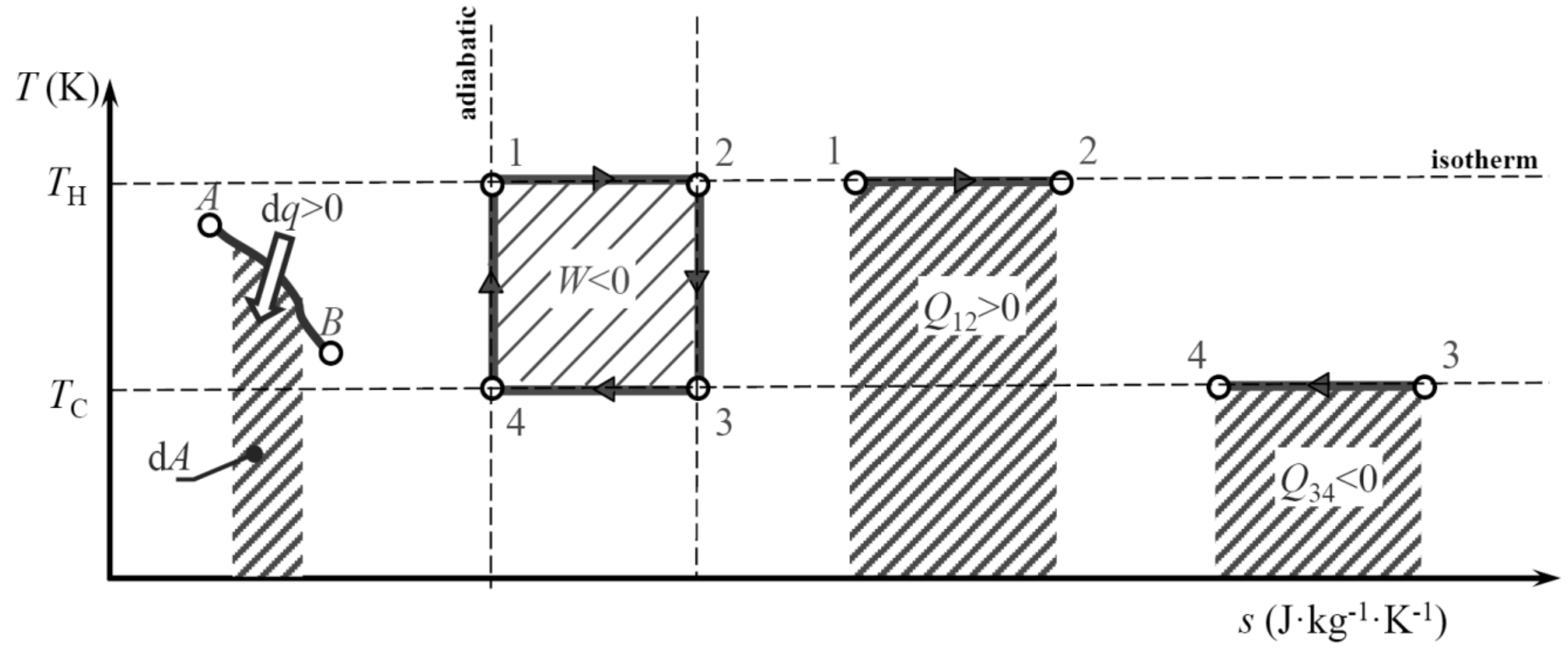

**Fig. 6:** Carnot cycle

The Carnot thermodynamic cycle (see Fig. 6) performs four successive reversible transforms:

- two isothermal processes for which reversibility means that the heat transfers occur under a very small temperature difference;
- two adiabatic processes for which reversibility leads to isentropic transformations.

This fully reversible cycle performs the maximum efficiency obtainable with a heat machine and thus serves as a reference for comparing efficiencies of heat machines, taking into account the temperatures of their heat reservoirs. The relative efficiencies of heat machines are thus defined as follows:

$$\text{engine } \eta_{\mathrm{r}} = \frac{\eta}{\eta_{\mathrm{C}}} = 1 - \frac{T_{\mathrm{C}} S^{\mathrm{i}}}{Q_{\mathrm{H}} \left(1 - T_{\mathrm{C}} / T_{\mathrm{H}}\right)}; \quad (26)$$

$$\text{heat pump } COP_{\mathrm{r\,Heating}} = \frac{COP_{\mathrm{Heating}}}{COP_{\mathrm{C\,Heating}}} = \frac{1 - T_{\mathrm{C}} / T_{\mathrm{H}}}{1 - T_{\mathrm{C}} / T_{\mathrm{H}} - T_{\mathrm{C}} S^{\mathrm{i}} / Q_{\mathrm{C}}}; \quad (27)$$

$$\text{refrigerator } COP_{\mathrm{r\,Cooling}} = \frac{COP_{\mathrm{Cooling}}}{COP_{\mathrm{C\,Cooling}}} = \frac{1 - T_{\mathrm{H}} / T_{\mathrm{C}}}{1 - T_{\mathrm{H}} / T_{\mathrm{C}} - T_{\mathrm{C}} S^{\mathrm{i}} / Q_{\mathrm{H}}}. \quad (28)$$

Carnot efficiencies of a heat pump and a refrigerator are plotted in Fig. 7 versus the ratio of the temperatures of their reservoirs. As *COP* tends to very small values as the cold temperature $T_{\mathrm{C}}$ tends to zero, the cryogenic refrigeration community often employs the inverse of the *COP* and speaks in terms of 'watt to supply per watt of produced cooling power'.

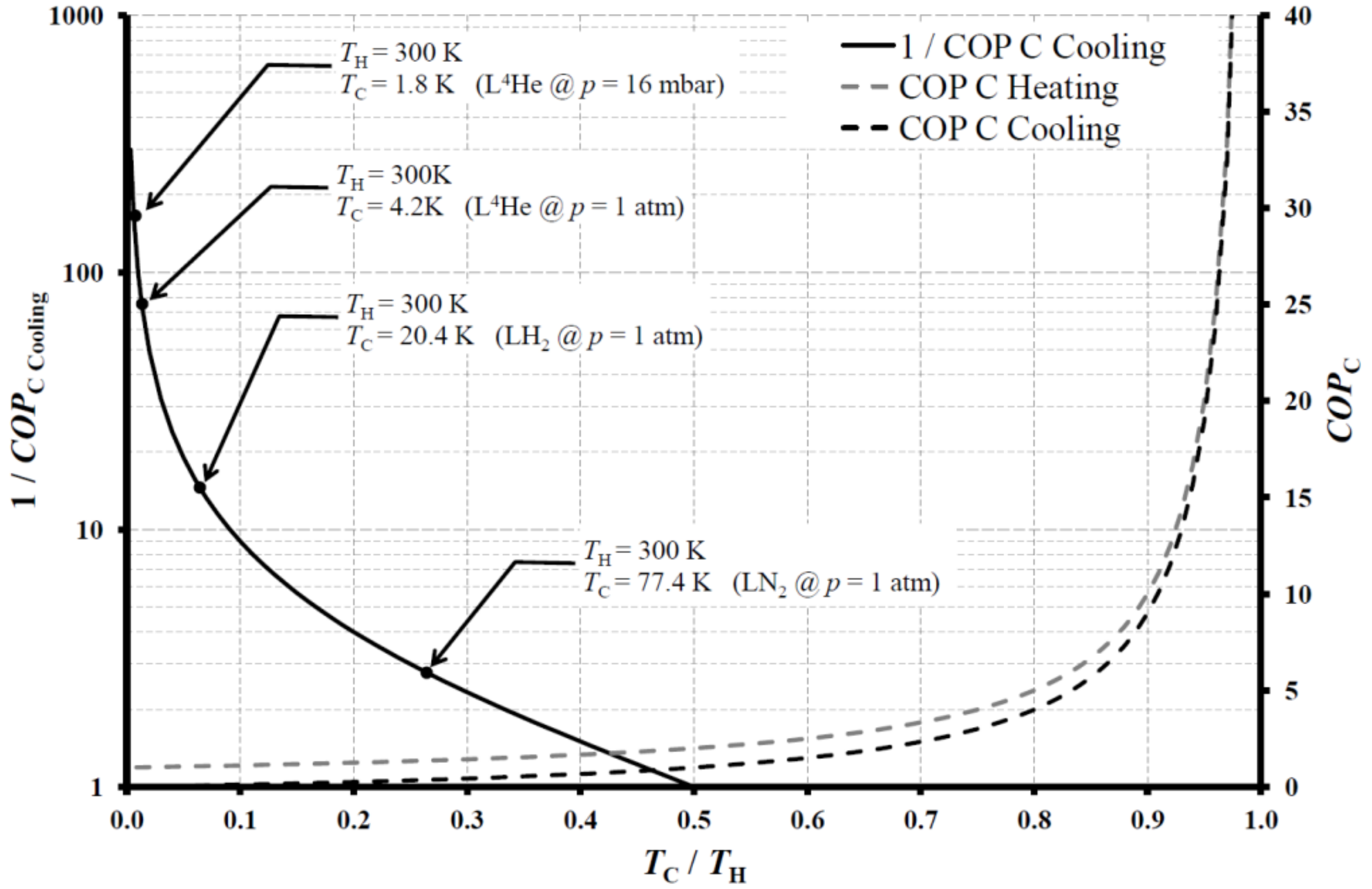

**Fig. 7:** *COP* values of a heat pump and a refrigerator based on the ideal Carnot cycle

The vapour compression cycle (see Fig. 8) achieved by our domestic refrigerators has a relatively good *COP* compared with the Carnot cycle. This intrinsically good performance comes from the fact that (i) the vaporization of a saturated liquid and the liquefaction of a saturated vapour are two isothermal process (heat is, however, transferred irreversibly); and (ii) the isenthalpic expansion of a saturated liquid is sensitively closed to an isentropic expansion. However, this cycle cannot be used in cryogenic applications; it relies on the phase change of a working fluid, and thus the temperatures of the heat sources cannot be very different. Even when implemented in cascade, the continuity of the process towards a low temperature cannot be insured as the existence of critical fluids (for which a liquid–gas phase transition is possible) does not cover continuously the domain of cryogenic temperature.

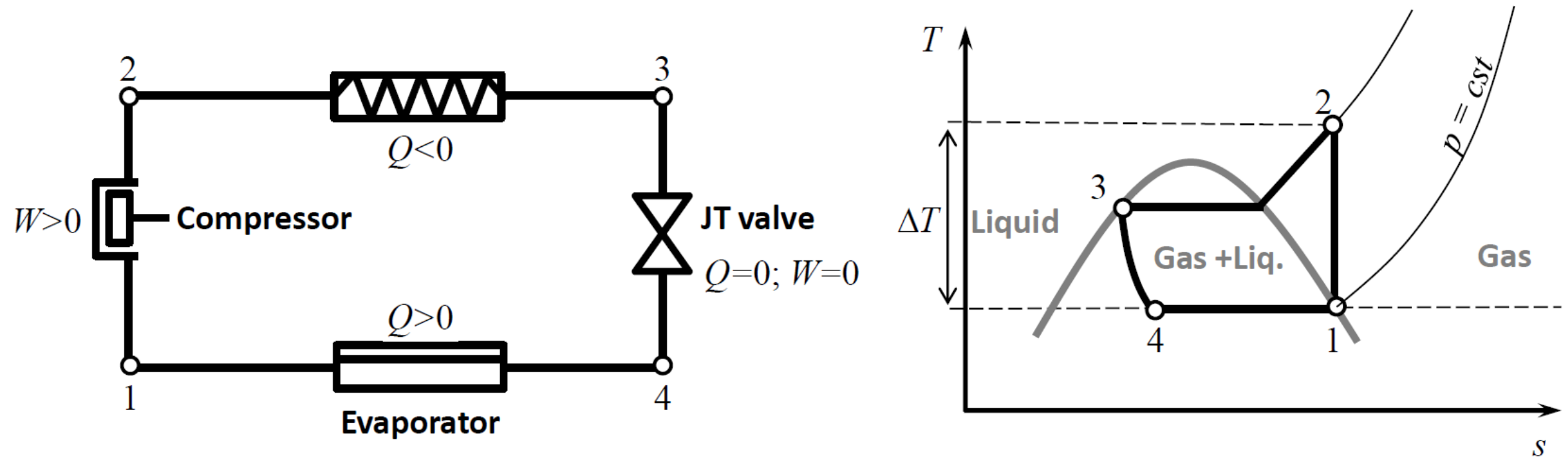

**Fig. 8:** Vapour compression heat machine

## 13 Maximum available work

Considering an isothermal ($T = cst$) and reversible thermodynamic process, $dU = \delta W + T\ \mathrm{d}S$ and thus $\delta W = \mathrm{d}U - T\ \mathrm{d}S = \mathrm{d}F$. The work produced by a system ($\delta W < 0$) is equal to the reduction of its free energy. For the reversible process, this work is the maximum that can be provided. Similarly, for an open system, $\mathrm{d}H = \delta W_{\mathrm{shaft}} + T\ \mathrm{d}S$ and thus $\delta W_{\mathrm{shaft}} = \mathrm{d}H - T\ \mathrm{d}S = \mathrm{d}G$. The maximum produced work (other than due to the external pressure forces) is equal to the reduction of the Gibbs free energy. In these cases, all the energy is free to perform useful work because there is no entropy production due to irreversibilities. Sources of entropy production come from heat transfers (due to temperature difference), friction (due to the movements of solid or fluid components), dissipative structure induced by fluid motions, matter diffusion, electric resistance, etc.

Heat and work are not equivalent: they do not have the same 'thermodynamic grade'. Whereas mechanical or electric work can be integrally converted into heat, integrally converting heat into work is impossible according to the second law. Moreover, transfers are generally irreversible, which implies that thermodynamic processes have a direction (heat flows from a hot to a cold body, mass transfers from high to low pressures, etc.). The *exergy* allows us to 'rank' energies by involving the concept of 'usable' or 'available' energy: it is the maximum useful work that brings a system into equilibrium *with its surroundings* at $T_{EXT}$ (generally the room temperature). Exergy is defined as

$$Ex = U - U_{EXT} + p_{EXT}(V - V_{EXT}) - T_{EXT}(S - S_{EXT}) \ , \tag{29}$$

such that it tends to zero when the energy $U$, the volume $V$, and the entropy $S$ of the system tend to those of the surroundings: $U_{EXT}$, $V_{EXT}$, and $S_{EXT}$. *Exergy is a state function of the system relative to its surroundings, not of the system only.*

Considering a closed system in thermal and pressure contact with its surroundings, we can deduce from the first and second law that

$$\Delta(U + p_{EXT}V - T_{EXT}S) = W_{\mathrm{u}} - T_{EXT}S^{\mathrm{i}} ,$$

where $W_{\mathrm{u}}$ is the useful work produced by the system (other than the work due to pressure forces). This work is maximum for $S^{\mathrm{i}} = 0$, and thus

$$W_{\mathrm{u}} \leq \Delta(U + p_{EXT}V - T_{EXT}S) = \Delta Ex \, . \tag{30}$$

The maximum produced work is equal to the reduction of the exergy $\Delta Ex$. If we define the Gibbs free energy of our system by imposing $T = T_{EXT}$ and $p = p_{EXT}$, we see that $\Delta Ex = \Delta G$. We can thus determine the exergy by use of the thermodynamic potentials:

$$\mathrm{d}Ex = \mathrm{d}H - T_{EXT}\ \mathrm{d}S \ ; \ \mathrm{d}Ex = \mathrm{d}F - p_{EXT}\ \mathrm{d}V \ ; \ \mathrm{d}Ex = \mathrm{d}G \, .$$

From the previous considerations, and in view of the energy and entropic balances, we can define the exergetic balance of an open system, noting that $m_i \cdot ex_i = Ex_i$ :

$$\frac{\mathrm{d}Ex}{\mathrm{d}t} = \frac{\mathrm{d}}{\mathrm{d}t}(U + p_{EXT}V - T_{EXT}S) = \sum_{i\in\text{inlet}} \dot{m}_i \cdot ex_i - \sum_{i\in\text{outlet}} \dot{m}_i \cdot ex_i + \dot{W}_\mathrm{u} + \sum_i \dot{Q}_i \left(1 - \frac{T_{EXT}}{T_i}\right) - T_{EXT}\dot{S}^\mathrm{i} . \quad (31)$$

The term $-T_{EXT}S^\mathrm{i}$ represents a destruction of exergy.

For a stationary state and a closed system, it reduces to

$$\dot{W}_\mathrm{u} + \sum_i \dot{Q}_i \left(1 - \frac{T_{EXT}}{T_i}\right) - T_{EXT}\dot{S}^\mathrm{i} = 0 . \quad (32)$$

The exergetic balance of one of the heat machines we considered in Section 12 is thus

$$\dot{W}_\mathrm{u} + \dot{Q}_\mathrm{H}\left(1 - \frac{T_{EXT}}{T_\mathrm{H}}\right) + \dot{Q}_\mathrm{C}\left(1 - \frac{T_{EXT}}{T_\mathrm{C}}\right) - T_{EXT}\dot{S}^\mathrm{i} = 0 .$$

Then, we can deduce that, for an engine, the higher the temperature of the supplied heat, the larger the work it can potentially produce.

## 14 States of matter and phase changes

We have already presented two equations of state for a compressible fluid: the ideal gas law and the van der Waals law. A pure substance can indeed be solid, liquid, or gaseous, and its state at thermodynamic equilibrium depends on the two intensive parameters $(p,T)$, as shown in the $p-T$ phase diagram of Fig. 9. Two phases coexist in equilibrium along each solid line. The three curves intercept at the triple point J, where the three phases coexist.

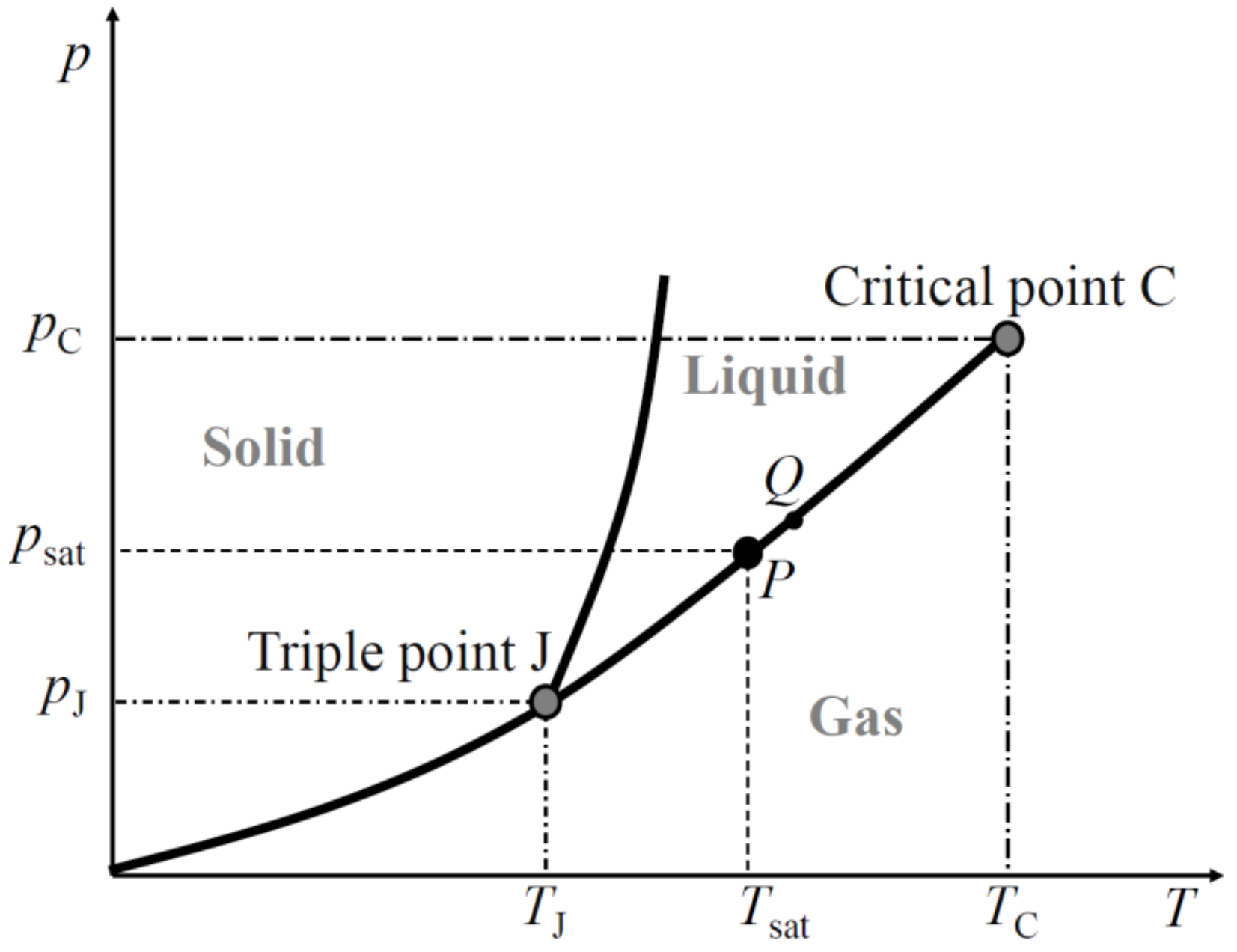

**Fig. 9:** $p-T$ phase diagram

The Gibbs phase rule gives the number of degrees of freedom $D$ (number of independent intensive variables) of a system of mixed substances. In the absence of a chemical reaction or external constraint, the variance of the system depends on the number of (chemically independent) constituents $c$ and the number of phases $\phi$ :

$$D = c + 2 - \phi \ . \tag{33}$$

The variance of a monophasic (in a single phase) pure substance in thermodynamic equilibrium is $D = 2$; a full description of the system requires both pressure and temperature to be set. At the *triple point* J, where the substance is at equilibrium in three phases, $D = 0$. For a pure substance, this unique point is defined by a unique couple $(p_J, T_J)$. Triple points of different pure substances are used as metrological references for the temperature scale and for the calibration of reference temperature sensors. When two phases coexist at equilibrium, the variance is equal to 1. Thus, setting $p$ or $T$ variable will fix the other one. This is the case at point $P$ (placed on the *saturated curve* $[\mathrm{JC}]$) on Fig. 9: the gas (or vapour) and liquid phases coexist in equilibrium. The condition for this equilibrium is that the free Gibbs energy $G$ tends to a minimum. Consider a small quantity of matter $\delta m$ that transfers from the liquid to the gas phase, then the change in the total free Gibbs energy is $\delta m(g_{\mathrm{gas}} - g_{\mathrm{liq}})$, which is minimum for $g_{\mathrm{gas}} = g_{\mathrm{liq}}$. If we now consider a neighbouring point $Q$ on the saturated liquid–vapour curve, we can write, per mass unit:

for the gaseous phase $g_{\mathrm{gas}}(Q) = g_{\mathrm{gas}}(P) + \left(\dfrac{\partial g_{\mathrm{gas}}}{\partial p}\right)_T \mathrm{d}p + \left(\dfrac{\partial g_{\mathrm{gas}}}{\partial T}\right)_p \mathrm{d}T = g_{\mathrm{gas}}(P) + v_{\mathrm{gas}}\, \mathrm{d}p - s_{\mathrm{gas}}\, \mathrm{d}T$ ,

for the liquid phase $g_{\mathrm{liq}}(Q) = g_{\mathrm{liq}}(P) + \left(\dfrac{\partial g_{\mathrm{liq}}}{\partial p}\right)_T \mathrm{d}p + \left(\dfrac{\partial g_{\mathrm{liq}}}{\partial T}\right)_p \mathrm{d}T = g_{\mathrm{liq}}(P) + v_{\mathrm{liq}}\, \mathrm{d}p - s_{\mathrm{liq}}\, \mathrm{d}T$ ,

which yields, by subtraction,

$$0 = (v_{\mathrm{gas}} - v_{\mathrm{liq}})\mathrm{d}p - (s_{\mathrm{gas}} - s_{\mathrm{liq}})\mathrm{d}T \ ,$$

where $v_{\mathrm{gas}}$ and $v_{\mathrm{liq}}$ (unit: $\mathrm{m^3 \cdot kg^{-1}}$) are the specific volumes (the reciprocals of the mass densities) of the gas and liquid phases.

The slope of the saturated vapour tension curve is thus given by the *Clausius–Clapeyron equation*:

$$\frac{\mathrm{d}p}{\mathrm{d}T} = \frac{(s_{\mathrm{gas}} - s_{\mathrm{liq}})}{(v_{\mathrm{gas}} - v_{\mathrm{liq}})} = \frac{l_{\mathrm{liq}\to\mathrm{vap}}}{T(v_{\mathrm{gas}} - v_{\mathrm{liq}})}, \tag{34}$$

where $l_{\mathrm{liq}\to\mathrm{vap}} = T(s_{\mathrm{gas}} - s_{\mathrm{liq}})$ ($>0$; unit: $\mathrm{J \cdot kg^{-1}}$) is the specific *latent heat* for the liquid → gas transition (vaporization). It is the quantity of heat required to transform, at constant temperature and pressure, a unit mass of liquid into vapour. We can express this specific latent heat by use of the enthalpy difference per mass unit between the vapour and the liquid. Indeed, during the reversible phase transition process,

$$l_{\mathrm{liq}\to\mathrm{vap}} = q = \int_{\mathrm{liq}}^{\mathrm{vap}} [\mathrm{d}u + p\, \mathrm{d}v] = \int_{\mathrm{liq}}^{\mathrm{vap}} [\mathrm{d}u + \mathrm{d}(pv)] = \int_{\mathrm{liq}}^{\mathrm{vap}} \mathrm{d}h = h_{\mathrm{vap}} - h_{\mathrm{liq}} \ . \tag{35}$$

Quantification of the vaporization latent heat from a $T - s$ diagram is shown in Fig. 10. It can be deduced that $l_{\mathrm{liq}\to\mathrm{vap}}$ tends to zero as we approach the critical point. Also, $l_{\mathrm{liq}\to\mathrm{vap}}$ can be expressed as a function of $T$ by neglecting $v_{\mathrm{liq}}$ (as $v_{\mathrm{gas}} >> v_{\mathrm{liq}}$) and using an equation of state to express $\mathrm{d}p/\mathrm{d}T$.

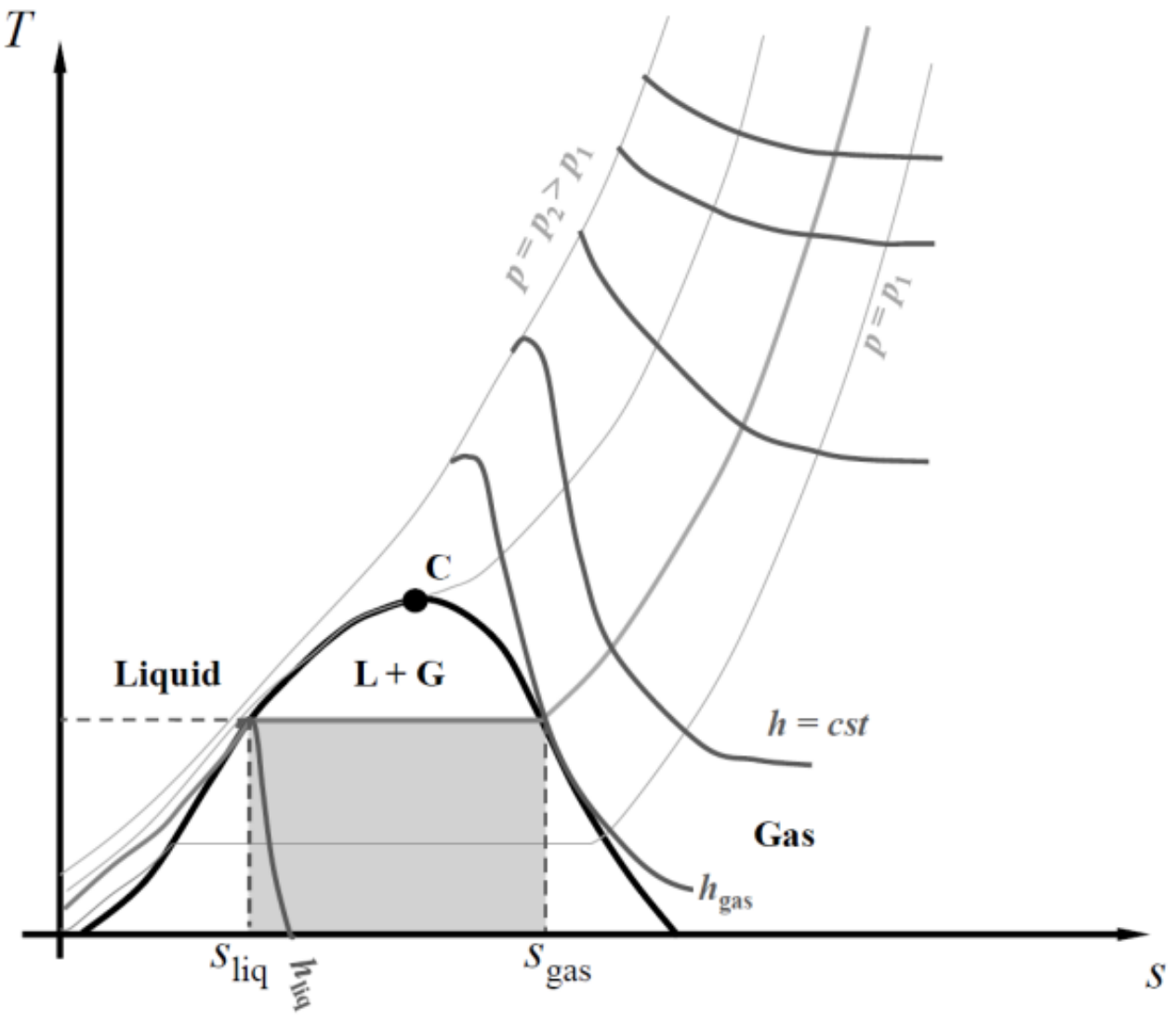

**Fig. 10:** Evaluation of the latent heat of the liquid–gas transition in a $T - s$ diagram

The same argument can be applied to solid–liquid and solid–gas phase transitions defining fusion (melting) and sublimation (deposition) latent heats, respectively. These phase transitions are characterized by a discontinuity in the first derivatives of the Gibbs free energy $S = -(\partial G / \partial T)_p$: at equilibrium, the entropy of the liquid phase is not the same as that of the gaseous phase. Thus, these phase transitions are *first-order transitions*.

At the critical point ($T = T_C; p = p_C$), however, $v_{gas} = v_{liq}$ and $\mathrm{d}p / \mathrm{d}T$ is finite so the Clausius–Clapeyron equation states that $l_{liq \to vap} = 0 = (s_{gas} - s_{liq})$. No latent heat is involved during the transition and $S = -(\partial G / \partial T)_p$ is continuous. But second derivatives of Gibbs free energy are not continuous, and the transition is of second order.

## 15 Thermodynamics of magnetic systems: type I superconductor phase transition

In a vacuum, the magnetic field $\hat{\mathbf{B}}$ (unit: T ≡ V·s·m$^{-2}$ ≡ N·A$^{-1}$·m$^{-1}$) is proportional to the excitation (or applied) magnetic field $\hat{\mathbf{H}}$ (unit: V·s·A$^{-1}$·m$^{-1}$ ≡ A·m$^{-1}$): $\hat{\mathbf{B}} = \mu_0 \hat{\mathbf{H}}$. The proportional factor is the permeability of free space, $\mu_0 = 4\pi \cdot 10^{-7}$ (N·A$^{-2}$). In a material, $\hat{\mathbf{B}} = \mu_0(\hat{\mathbf{H}} + \hat{\mathbf{M}})$, where $\hat{\mathbf{M}} = \chi \hat{\mathbf{H}}$ (unit: A·m$^{-1}$) is the magnetization, which represents how strongly a region of material is magnetized by the application of $\hat{\mathbf{H}}$; $\chi$ is the magnetic susceptibility. Thus, $\hat{\mathbf{B}} = \mu_0(1 + \chi)\hat{\mathbf{H}} = \mu_0 \mu_r \hat{\mathbf{H}}$, where $\mu_r$ is the relative magnetic permeability.

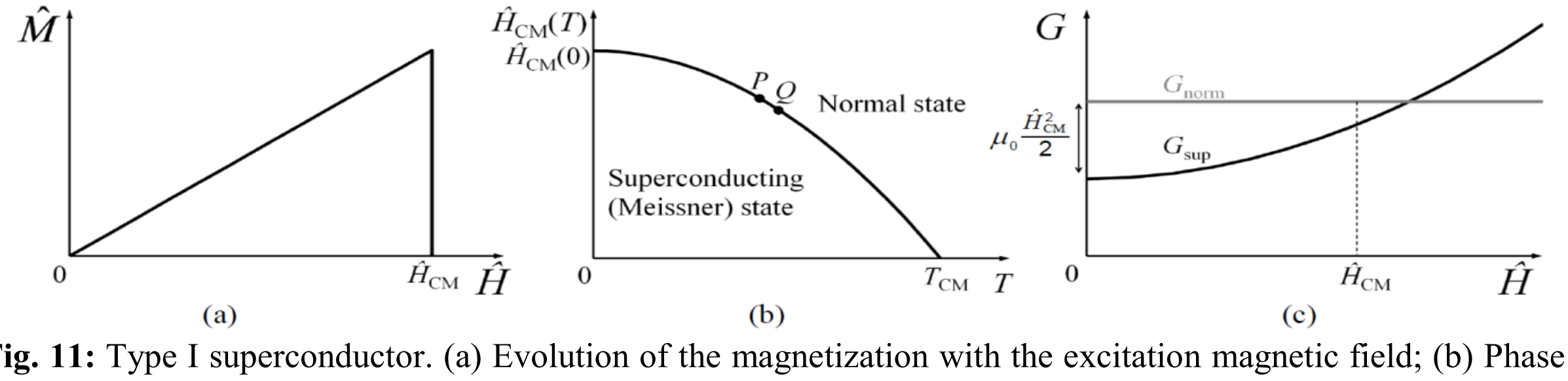

**Fig. 11:** Type I superconductor. (a) Evolution of the magnetization with the excitation magnetic field; (b) Phase diagram; (c) Evolution of the Gibbs free energy with the excitation magnetic field below the critical temperature.

We consider a long cylinder sample of type I superconductor placed in a solenoid producing the excitation magnetic field $\hat{H} = n \cdot i$. Note that in the Meissner state, i.e. below a critical temperature $T_{\mathrm{CM}}$ and below a critical magnetic field $\hat{\mathbf{H}}_{\mathrm{CM}}(T)$, type I superconductors are perfect diamagnets, with $\chi = -1$, $B = 0$, and $\hat{\mathbf{M}} = -\hat{\mathbf{H}}$ (see Fig. 11(a)). We assume that the sample has a uniform magnetization parallel to the field. If we increase the external field from $\hat{H}$ to $\hat{H} + \mathrm{d}\hat{H}$, it induces a change $d\hat{B}$ of $\hat{B}$ and $\mathrm{d}\hat{M}$ of $\hat{M}$. The resulting reversible magnetic work produced per unit volume (unit: J·m$^{-3}$) is given by

$$\mathrm{d}W_{TOT} = \hat{H} \cdot \mathrm{d}\hat{B} = \mu_0(\hat{H} \cdot \mathrm{d}\hat{H} + \hat{H} \cdot \mathrm{d}\hat{M}).$$

The first term is the energy of the magnetic field produced by the solenoid in the absence of the material. The second term is the energy supplied to the sample: $\mathrm{d}W = \mu_0 \hat{H} \cdot \mathrm{d}\hat{M}$. The change of internal energy of the material for this reversible process is then $\mathrm{d}U = T\,\mathrm{d}S + \mu_0 \hat{H} \cdot \mathrm{d}\hat{M}$. Considering $\hat{H} \equiv -p$ and $\hat{M} \equiv V$, we see that it is analogous with $\mathrm{d}U = T\,\mathrm{d}S - p\,\mathrm{d}V$. Similar to the liquid–gas phase transition we studied at fixed $p$ and $T$, this phase transition from the superconducting to the normal state at fixed $\hat{H}$ and $T$ can be studied using the Gibbs free energy (per unit volume):

$$\hat{G} = U - TS - \mu_0 \hat{H}\hat{M}. \qquad (36)$$

At constant sample volume $V$, the condition for coexistence of the two phases (normal and superconducting) at equilibrium along the critical curve $\hat{H}_{\mathrm{CM}}(T)$ is that the total free Gibbs energy tends to a minimum, and so Gibbs free energy is equal in the two phases: $\hat{G}_{\mathrm{norm}} = \hat{G}_{\mathrm{sup}}$. Below the critical temperature ($T < T_{\mathrm{CM}}$), as for point $P$ in Fig. 11(b), equilibrium is reached for $\hat{H} = \hat{H}_{\mathrm{CM}}(T)$, and thus $\hat{G}_{\mathrm{norm}}(T, \hat{H}_{\mathrm{CM}}) = \hat{G}_{\mathrm{sup}}(T, \hat{H}_{\mathrm{CM}})$. For a neighbouring point $Q$ on the critical curve, $\hat{G}_{\mathrm{norm}}(T+\mathrm{d}T, \hat{H}_{\mathrm{CM}}+\mathrm{d}\hat{H}_{\mathrm{CM}}) = \hat{G}_{\mathrm{sup}}(T+\mathrm{d}T, \hat{H}_{\mathrm{CM}}+\mathrm{d}\hat{H}_{\mathrm{CM}})$, and by subtraction $\mathrm{d}\hat{G}_{\mathrm{norm}} = \mathrm{d}\hat{G}_{\mathrm{sup}}$. Taking into account that (per unit volume)

$$\mathrm{d}\hat{G} = \mathrm{d}U - T\,\mathrm{d}S - S\,\mathrm{d}T - \mu_0 \hat{H} \cdot \mathrm{d}\hat{M} - \mu_0 \hat{M} \cdot \mathrm{d}\hat{H} = -S\,\mathrm{d}T - \mu_0 \hat{M} \cdot \mathrm{d}\hat{H},$$

and that the magnetization in the normal state is negligible ($\hat{M}_{\mathrm{norm}} \ll 1$), we obtain

$$(S_{\mathrm{norm}} - S_{\mathrm{sup}}) = -V \cdot \mu_0 \cdot \hat{H}_{\mathrm{CM}} \cdot \frac{\mathrm{d}\hat{H}_{\mathrm{CM}}}{\mathrm{d}T} = -V \frac{1}{2} \frac{\mathrm{d}\hat{H}_{\mathrm{CM}}^2}{\mathrm{d}T} \quad \text{(unit: J)}. \qquad (37)$$

Experimentally, $\hat{H}_{\mathrm{CM}}(T) = \hat{H}_{\mathrm{CM}}(0) - \hat{H}_{\mathrm{CM}}(0) \cdot (T / T_{\mathrm{CM}})^2$ (see Fig. 11(b)), $\mathrm{d}\hat{H}_{\mathrm{CM}}^2 / \mathrm{d}T \leq 0$, and thus $S_{\mathrm{norm}} \geq S_{\mathrm{sup}}$; in the Meissner state, the material has a lower entropy due to the Cooper pairing of electrons in a single quantum state. Integrating the Gibbs free energy along an increasing $\hat{H}$, we find that for the phase in the superconducting state ($\hat{M} = -\hat{H}$)

$$\hat{G}_{\mathrm{sup}}(\hat{H}, T) = \hat{G}_{\mathrm{sup}}(0, T) - \mu_0 \int_0^{\hat{H}} M\,\mathrm{d}H = \hat{G}_{\mathrm{sup}}(0, T) + \frac{1}{2}\mu_0 \hat{H}^2. \qquad (38)$$

As $\hat{M}_{\mathrm{norm}} \ll 1$, the Gibbs free energy of the normal state phase is not dependent on $\hat{H}$. The Gibbs free energies (per unit volume) of the normal and superconducting phases are plotted in Fig. 11(c) as a function of the excitation magnetic field. We see that for $T < T_{\mathrm{CM}}$, $\hat{G}_{\mathrm{sup}} < \hat{G}_{\mathrm{norm}}$: the superconducting state has a lower free energy. However, as $T > T_{\mathrm{CM}}$, $\hat{G}_{\mathrm{sup}} > \hat{G}_{\mathrm{norm}}$, and transition to the normal state occurs, the phase of minimum free enthalpy.

The entropy difference is related to the latent heat of the phase transition:

$$L_{\mathrm{norm}\to\mathrm{sup}} = (S_{\mathrm{norm}} - S_{\mathrm{sup}})T = -V\mu_0 T \cdot \hat{H}_{\mathrm{CM}} \frac{\mathrm{d}\hat{H}_{\mathrm{CM}}}{\mathrm{d}T} = 2V\mu_0 \hat{H}_{\mathrm{CM}}^2(0) \frac{T^2}{T_{\mathrm{CM}}^2}\left(1 - \frac{T^2}{T_{\mathrm{CM}}^2}\right) \quad \text{(unit: J)}. \qquad (39)$$

As $T < T_{\mathrm{CM}}$, $L_{\mathrm{norm}\to\mathrm{sup}} > 0$; during the transition from the superconducting to the normal state due to the magnetic field, the material absorbs a quantity of heat which is provided by this latent heat. This is also a first-order transition as the first derivative of the Gibbs free energy $S = -(\partial G / \partial T)_p$ is not continuous from one phase to another. At $T = T_{\mathrm{CM}}$, $L_{\mathrm{norm}\to\mathrm{sup}} = 0$ and the entropy is continuous.

However, the heat capacity $C = \delta Q / \mathrm{d}T = T\,\mathrm{d}S / \mathrm{d}T = -T(\partial^2 G / \partial T^2)_p$, a second derivative of the Gibbs energy, is not continuous:

$$(C_{\mathrm{norm}} - C_{\mathrm{sup}}) = -4\mu_0 V \hat{H}^2_{\mathrm{CM}} / T_{\mathrm{CM}} . \tag{40}$$

Therefore at $T = T_{\mathrm{CM}}$ the superconducting → normal transition is a second-order transition.

## 16 Second-order phase transitions

The liquid–gas transition of a fluid and the superconducting–normal state of a type I superconductor are first-order transitions, except when the temperature reaches the critical temperature: no latent heat is involved and they become of second-order transitions as the second derivatives of the Gibbs free energy are not discontinuous.

Continuity of the first derivatives of the Gibbs free energy (per unit of mass) during a second-order transition between phases I and II implies that $\mathrm{d}s_{\mathrm{I}} = \mathrm{d}s_{\mathrm{II}}$ and $\mathrm{d}v_{\mathrm{I}} = \mathrm{d}v_{\mathrm{II}}$. Using Eqs. (15), (16) and (20) we find that

$$\mathrm{d}s_i = \left(\frac{\partial s_i}{\partial T}\right)_p \mathrm{d}T + \left(\frac{\partial s_i}{\partial p}\right)_T \mathrm{d}p = \frac{c_{pi}}{T}\mathrm{d}T - \left(\frac{\partial v_i}{\partial T}\right)_p \mathrm{d}T = \frac{c_{pi}}{T}\mathrm{d}T - v_i\alpha_i\,\mathrm{d}p .$$

Similarly, using Eq. (20) and introducing the coefficient of isothermal compressibility $\chi_T = -1/V(\delta V/\delta p)_T$, we obtain

$$\mathrm{d}v_i = \left(\frac{\partial v_i}{\partial T}\right)_p \mathrm{d}T + \left(\frac{\partial v_i}{\partial p}\right)_T \mathrm{d}p = v_i\alpha_i\,\mathrm{d}T - v_i\chi_{Ti}\,\mathrm{d}p .$$

Hence, from the equality of the derivatives on the coexistence curve, and as there is no change in volume during the transition, we obtain, for the Ehrenfest formulae,

$$\frac{\mathrm{d}p}{\mathrm{d}T} = \frac{c_{p\mathrm{I}} - c_{p\mathrm{II}}}{Tv(\alpha_{\mathrm{I}} - \alpha_{\mathrm{II}})} \quad \text{and} \quad \frac{\mathrm{d}p}{\mathrm{d}T} = \frac{\alpha_{\mathrm{I}} - \alpha_{\mathrm{II}}}{\chi_{T\mathrm{I}} - \chi_{T\mathrm{II}}} . \tag{41}$$

An example of a second-order transition is the normal–superfluid transition of liquid helium-4. The equilibrium curve of these two phases in a $p - T$ diagram is called the *lambda line*. This name originates from the shape of the heat capacity curve plotted versus temperature, which has the form of the Greek letter lambda near the transition, as can be seen in Fig. 12(a). The *lambda point* is the triple point where the normal phase, the superfluid liquid phase, and the vapour phase coexist. As $T$ diminishes to $T_\lambda$, the heat capacity of He-I (normal state) increases. At $T = T_\lambda$, the heat capacity is discontinuous, as shown in Fig. 12(b). Below $T_\lambda$, the heat capacity reaches a maximum and decreases as $T$ diminishes.

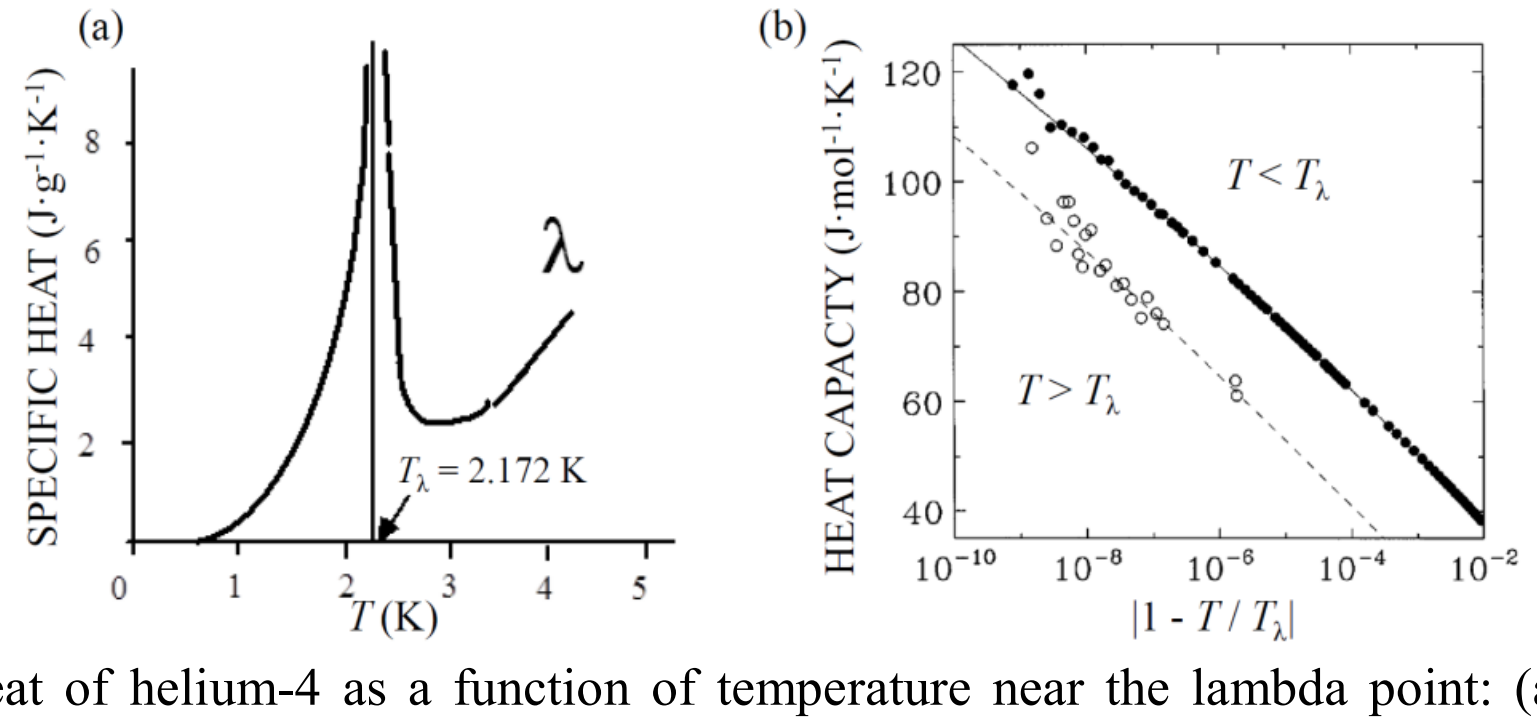

**Fig. 12:** Specific heat of helium-4 as a function of temperature near the lambda point: (a) large temperature scale; (b) very close to $T_\lambda$ (from [1]).

It should be noted that higher-order phase transitions can be found in nature, even though they are less abundant than first- or second-order ones. Their order is defined as the minimum integer $i$ such that $(\partial^i G / \partial x^i)_{\mathrm{I}} \neq (\partial^i G / \partial x^i)_{\mathrm{II}}$.

## References

[1] J.A. Lipa *et al.*, *Phys. Rev. Lett.* **76** (1995) 944.

## Appendix A: Thermodynamic processes for an ideal gas

**Table A.1:** Summary of the properties of some reversible processes involving an ideal gas

| **Process** | **Isochoric** | **Isobaric** | **Isothermal** | **Adiabatic** | **Polytropic** |
|---|---|---|---|---|---|
| Fixed parameter | $V = cst$ | $p = cst$ | $T = cst$ | $\delta q = 0$ | |
| $p$–$v$ diagram | 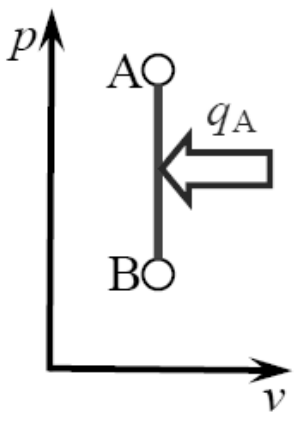 | 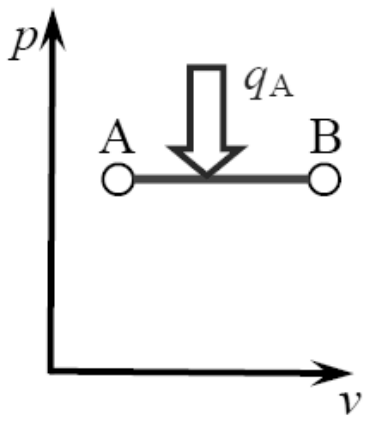 | 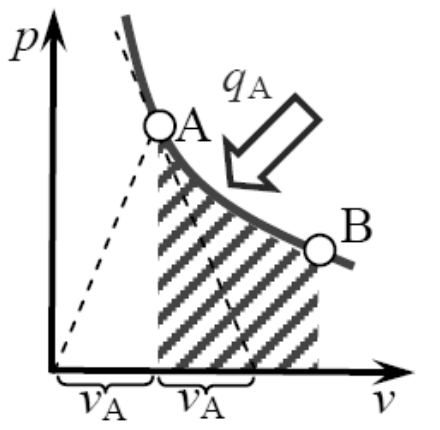 | 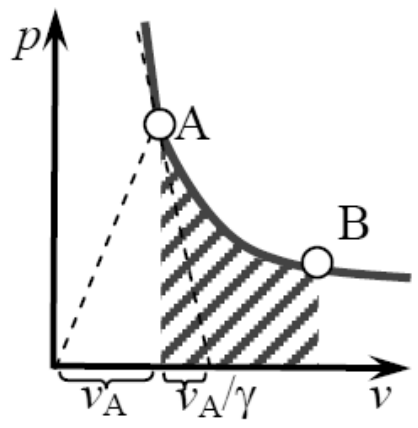 | 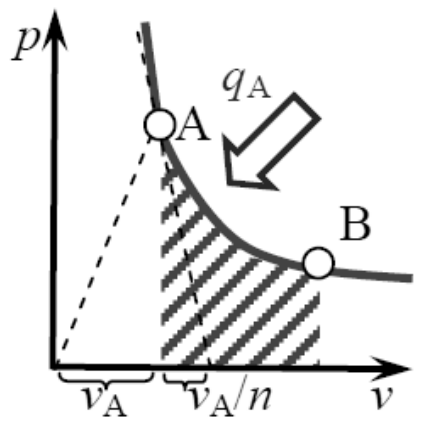 |
| Relationship between state variables | $\frac{p_B}{p_A} = \frac{T_B}{T_A}$ Charles law | $\frac{v_B}{v_A} = \frac{T_B}{T_A}$ Gay–Lussac law | $p_A v_A = p_B v_B = nRT_A$ Mariotte law | $p_A v_A^\gamma = p_B v_B^\gamma$ Laplace law $\gamma = c_p / c_V$ | $p_A v_A^z = p_B v_B^z$ $1 < z < \gamma$ |
| $\Delta U = U_B - U_A$ (J) | $nc_V(T_B - T_A)$ | $nc_V(T_B - T_A)$ | $nc_V(T_B - T_A) = 0$ | $nc_V(T_B - T_A)$ | $nc_V(T_B - T_A)$ |
| $\Delta H = H_B - H_A$ (J) | $nc_p(T_B - T_A)$ | $nc_p(T_B - T_A)$ | $nc_p(T_B - T_A) = 0$ | $nc_p(T_B - T_A)$ | $nc_p(T_B - T_A)$ |
| Heat $Q_{AB}$ (J) | $\Delta U$ | $\Delta H$ | $nRT \ln \frac{V_B}{V_A}$ | 0 | $nc_v \frac{z-\gamma}{z-1}(T_B - T_A)$ |
| Work $W_{AB}$ (J) | 0 | $-p(V_B - V_A) = nR(T_B - T_A)$ | $-nRT_A \ln \frac{V_B}{V_A}$ | $\Delta U$ | $nc_v \frac{\gamma-1}{z-1}(T_B - T_A)$ |
| $\Delta S = S_B - S_A$ ($\mathrm{J \cdot K^{-1}}$) | $nc_V \ln \frac{T_B}{T_A}$ | $c_p \ln \frac{T_B}{T_A} = c_p \ln \frac{V_B}{V_A}$ | $nR \ln \frac{V_B}{V_A}$ | 0 | $nR \frac{\gamma - n}{\gamma - 1} \ln \frac{V_B}{V_A}$ |